\documentclass[12pt]{article}
\usepackage{amsmath,epsf,amssymb,enumerate}
\newtheorem{theorem}{Theorem}

\newif\iffigs\figstrue

\DeclareFontFamily{U}{rsf}{}
\DeclareFontShape{U}{rsf}{m}{n}{
  <5> <6> rsfs5 <7> <8> <9> rsfs7 <10-> rsfs10}{}
\DeclareMathAlphabet\Scr{U}{rsf}{m}{n}

\def\pplogo{\vbox{\kern-\headheight\kern -29pt
\halign{##&##\hfil\cr&{
\ppnumber}\cr\rule{0pt}{2.5ex}&\ppdate\cr}
}}
\makeatletter
\def\ps@firstpage{\ps@empty \def\@oddhead{\hss\pplogo}%
  \let\@evenhead\@oddhead 
}
\def\maketitle{\par
 \begingroup
 \def\thefootnote{\fnsymbol{footnote}}
 \def\@makefnmark{\hbox{$^{\@thefnmark}$\hss}}
 \if@twocolumn
 \twocolumn[\@maketitle]
 \else \newpage
 \global\@topnum\z@ \@maketitle \fi\thispagestyle{firstpage}\@thanks
 \endgroup
 \setcounter{footnote}{0}
 \let\maketitle\relax
 \let\@maketitle\relax
 \gdef\@thanks{}\gdef\@author{}\gdef\@title{}\let\thanks\relax}
\makeatother

\def\C{{\mathbb C}}
\def\P{{\mathbb P}}
\def\Q{{\mathbb Q}}
\def\R{{\mathbb R}}
\def\Z{{\mathbb Z}}

\def\Hom{\operatorname{Hom}}

\def\Jac{\operatorname{Jac}}

\def\GO{\operatorname{O{}}}
\def\SU{\operatorname{SU}}

\def\rank{\operatorname{rank}}
\def\Spin{\operatorname{Spin}}

\def\CY{Calabi--Yau}

\def\MW{Mordell--Weil}

\def\cM{{\Scr M}}

\def\cD{{\Scr D}}

\def\cG{{\Scr G}}

\def\cMc{{\hfuzz=100cm\hbox to 0pt{$\;\overline{\phantom{X}}$}\cM}}
\def\barcD{{\hfuzz=100cm\hbox to 0pt{$\;\overline{\phantom{X}}$}\cD}}
\def\ff#1#2{{\textstyle\frac{#1}{#2}}}

\def\spnh{\Spin(32)/\Z_2}

\def\HS#1{{\mathbb{F}}_{#1}}

\def\mf#1{\mathfrak{#1}}

\begin{document}
\setcounter{page}0
\title{\LARGE Aspects of the Hypermultiplet Moduli Space\\ in String
 Duality\\[10mm]} 
\author{
Paul S. Aspinwall\\[10mm]
\normalsize Center for Geometry and Theoretical Physics, \\
\normalsize Box 90318, \\
\normalsize Duke University, \\
\normalsize Durham, NC 27708-0318\\[10mm]
}
\def\ppnumber{\vbox{\baselineskip14pt\hbox{DUKE-CGTP-98-01}
\hbox{hep-th/9802194}}}
\def\ppdate{February 1998} \date{}

{\hfuzz=10cm\maketitle}

\def\Large{\large}
\def\LARGE{\large\bf}


\begin{abstract}
A type IIA string (or F-theory) compactified on a \CY\ threefold is
believed to be dual to a heterotic string on a K3 surface times a
2-torus (or on a K3 surface). We consider how the resulting moduli space
of hypermultiplets is identified between these two pictures in the
case of the $E_8\times E_8$ heterotic string. As
examples we discuss $\SU(2)$-bundles and $G_2$-bundles on the K3
surface and the case of point-like instantons.
We are lead to a rather beautiful identification between the integral
cohomology of the \CY\ threefold and some integral structures on the
heterotic side somewhat reminiscent of mirror symmetry.
We discuss the consequences for probing nonperturbative effects in the
both the type IIA string and the heterotic string.
\end{abstract}

\vfil\break

\def\HetG{\mathcal{G}_0}


\section{Introduction}    \label{s:int}

It has long been supposed that a type IIA string suitably compactified on a
\CY\ threefold is dual to a heterotic string suitably compactified on
a product of a K3 surface and a 2-torus \cite{KV:N=2}. The resulting
physics in four dimensions consists of a theory with $N=2$
supersymmetry. At least locally, the moduli space for these theories
is a product of a special K\"ahler manifold corresponding to moduli in
the vector supermultiplets, and a quaternionic K\"ahler manifold
corresponding to moduli in the hypermultiplets. The vector moduli
space has been fairly well understood for some time now (see, for
example, \cite{me:lK3} and references therein). The hypermultiplet
space has been somewhat more awkward to understand (although some
progress has been made, see for example \cite{BBS:5b}).

An almost equivalent problem arises in F-theory. If one compactifies
F-theory on a \CY\ threefold it can be dual to the heterotic string
compactified on a K3 surface. Our resulting theory is now an N=1
theory in six dimensions. There is still a quaternionic K\"ahler
moduli space corresponding to the moduli in the
hypermultiplets. Indeed, if this six-dimensional theory is
compactified on a 2-torus, we recover the $N=2$ theory in four
dimensions above with an unchanged hypermultiplet moduli 
space.\footnote{Except that a few parameters, such as the volume of
the K3 surface, are reinterpreted in the F-theory picture. We use the
type IIA language throughout this paper.}

Let us consider the type IIA string compactified on a particular \CY\
threefold, $X$. The hypermultiplet moduli space is then composed of
deformations of the complex structure of $X$, together with
Ramond-Ramond moduli living in the ``intermediate Jacobian'' of $X$,
and the dilaton and axion. On the other hand, in the heterotic string
picture the hypermultiplet moduli space consists of deformations of a
particular bundle on the K3 surface together with deformations of the
underlying K3 surface itself. The central question in understanding
the moduli space of hypermultiplets is to know exactly how to match
this data between the type IIA and heterotic picture. 

Thus somehow given a \CY\ threefold together with its intermediate
Jacobian, we should be able to ``derive'' some K3 surface with
a particular bundle. This is not a property of known classical
geometry. A very similar statement could be made concerning mirror
symmetry. In this case a type IIA string compactified on a \CY\
threefold $X$ is dual to a type IIB string compactified on the mirror
\CY\ threefold $Y$. Again one did not know classically how to
``derive'' $Y$ given $X$.

The key idea in mirror symmetry was to think in terms of large radius
limits and ``large complex structure'' limits. Similarly this paper
will dwell on the analogous boundary in the hypermultiplet moduli
space. This latter question is much more interesting than that of
mirror symmetry. Indeed this whole subject of exploring the moduli
space of hypermultiplets looks like a much richer version of the story
of mirror symmetry. 

The main purpose of the paper is to identify an integer structure on
the type IIA side, coming from $H^3(X,\Z)$, with an integer structure
on the heterotic side coming from $H^2(\mathrm{K3},\Z)$ and the
bundle data. This is a very useful handle for picturing how the moduli
space of hypermultiplets is mapped between the type IIA string and the
heterotic string.

In section \ref{s:sd} we will discuss the boundary of the
moduli space we need to study to remove all quantum effects. 
This includes a discussion of the $B$-field on the heterotic K3 surface.
In section \ref{s:bun} we will discuss the moduli space of bundles on a
K3 surface in the language of string duality. Much of this section
follows from the work of Friedman, Morgan and Witten \cite{FMW:F} but
we explain the construction in detail for a few examples. Finally in
section \ref{s:disc} we discuss the integral structure and some of its
consequences.


\section{The Stable Degeneration}    \label{s:sd}

\subsection{The geometry of the degeneration}  \label{ss:geom}

The type IIA string compactified on a \CY\ threefold, $X$, has the
following moduli:
\begin{enumerate}
\item The dilaton, which governs the string coupling, and the axion
coming from dualizing a two-form in four dimensions. Together these
form a complex field, $\Phi_{\text{IIA}}$. We assert that
$\Phi_{\text{IIA}}\to-\infty$ is the 
weakly-coupled limit of this type IIA string theory.
\item A metric which is determined by the complex structure of $X$ and
the cohomology class of its K\"ahler form. When all length scales in
$X$ are large with respect to the string tension, this metric is
Ricci-flat.
\item A $B$-field which takes values in $H^2(X,\R/\Z)$ (at least when
$X$ is large).
\item Ramond-Ramond fields, $R\in H^{\text{odd}}(X,\R/\Z)$.%
\footnote{It has been suggested \cite{HM:alg2} that the charges of BPS
states lie in $H^*(X,\Z/N)$ for some integer $N>1$. This may cause the
RR moduli to live in $H^{\text{odd}}(X,\R/(\Z/N))$. If this is the
case then some of the statements in this paper need to be modified to
take this isogeny into account.}
\end{enumerate}

The K\"ahler form modulus and the $B$-field pair up to form the
``complexified K\"ahler form''. The moduli space of vector moduli,
$\cM_V$,  is parametrized by this complexified K\"ahler form. There
are world-sheet instanton corrections to this moduli space however and
the complexified K\"ahler form is only a valid coordinate near the
large radius limit of $X$. We will also refer to these world-sheet
instantons as $\alpha'$ corrections as their scale is determined by
the string tension. $\cM_V$ is a special K\"ahler manifold (at least
where it is smooth away from any boundary points).

The other fields, namely $\Phi_{\text{IIA}}$, $R$, and the complex
structure of $X$ 
parameterize the hypermultiplet moduli space. Again this moduli space
is subject to corrections except that this time the corrections are
due to string coupling. The above parameters are only good coordinates
when we are near the limit of a weakly-coupled string,
$\Phi_{\text{IIA}}\to-\infty$.
The moduli space built from these parameters is a quaternionic
K\"ahler manifold and we denote it $\cM_H$.

Now let us consider the heterotic string. There are two types of
heterotic strings which we label by $\HetG$. $\HetG$ is either equal
to $\spnh$ or $(E_8\times E_8)\rtimes\Z_2$, where the latter case is
usually referred to as the $E_8\times E_8$ heterotic 
string.\footnote{The extra $\Z_2$ arises as the possibility of
exchanging the two $E_8$'s. It is essential to include this
possibility in order to understand duality between the two heterotic
strings correctly \cite{LSMT:w2,W:w2t}.}

The heterotic string compactified on a product of a K3 surface, $S_H$, and
a 2-torus, $E_H$, has the following moduli:
\begin{enumerate}
\item The dilaton, which governs the string coupling, and the axion
coming from dualizing a two-form in four dimensions. Together these
form a complex field, $\Phi_{\text{Het}}$. We assert that
$\Phi_{\text{Het}}\to-\infty$ is the 
weakly-coupled limit of this heterotic string theory.
\item A Ricci-flat metric on $S_H\times E_H$.
\item A $B$-field which takes values in $H^2(S_H\times E_H,\R/\Z)$ (at
least when $S_H$ and $E_H$ are large).
\item A $\HetG$-bundle on $S_H\times E_H$ with a connection satisfying
the Yang-Mills equation.
\end{enumerate}
In order to proceed further we need to make a restriction on the type
of this bundle. Let $\HetG \supset \mathcal{G}_S\times\mathcal{G}_E$ and
assume that the $\HetG$-bundle on $S_H\times E_H$ may be viewed as
a product of a $\mathcal{G}_S$-bundle on $S_H$ and a
$\mathcal{G}_E$-bundle on $E_H$.

The vector multiplet moduli space, $\cM_V$, can now be viewed as being
parametrized by $\Phi_{\text{Het}}$, the moduli of the
$\mathcal{G}_E$-bundle on $E_H$ and by the deformations of the metric
and $B$-field on $E_H$. This moduli space is subject to
corrections from the string coupling and identifications from
T-dualities for $E_H$. The above parameters are only
seen as good coordinates when $\Phi_{\text{Het}}\to-\infty$ and the
area of $E_H$ is large.

The hypermultiplet moduli space, $\cM_H$, can be viewed as being
parametrized by the moduli of the $\mathcal{G}_S$-bundle on $S_H$ and by
the deformations of the metric 
and $B$-field on $S_H$. The corrections to this moduli space are,
as yet, not fully understood but we will see that this picture is probably
prone to $\alpha'$ corrections. What is clear is that the above
parameters are good coordinates when the volume of $S_H$ is large.

The main purpose of this paper will be to try to match the type IIA
(or F-theory) picture of the moduli space $\cM_H$ with that of the
heterotic string. Going to the limits in which we remove the quantum
corrections, we may match our coordinate systems and obtain a map
between a \CY\ threefold, $X$ and a bundle on a K3 surface, $S_H$. In
order to do this we need to take $\Phi_{\text{IIA}}\to-\infty$ and
$S_H$ to have a large volume.

The first thing we should worry about are geometrical restrictions on
$X$ for the above duality to be possible. We want to impose the
conditions that 
\begin{enumerate}
\item We do have a heterotic string dual to the type IIA string on $X$
which can be recognized as such in a simple way.
\item The $\HetG$-bundle does indeed factorize nicely as a
$\mathcal{G}_S$-bundle and a $\mathcal{G}_E$-bundle as desired.
\end{enumerate}
It can be shown (see, for example, \cite{me:lK3}) that these
conditions amount to asking that $X$ has a dual fibration --- firstly
a K3-fibration $p:X\to B$ and as an elliptic fibration
$\pi_F:X\to\Sigma$. Here $B$ is $\P^1$, $\Sigma$ is a ruled rational
surface and $\pi_F$ has at least one section.

Now we concern ourselves with the question of how to go to the right
boundary in the moduli space to remove the quantum corrections.
First let us deal with $\Phi_{\text{IIA}}$. Let $T$ be a 2-torus in
$S_H$ which is a smoothly embedded elliptic curve for a suitable
choice of complex structure on $S_H$. Let $e_0$ be a 2-form which is
dual to the 2-cycle which is Poincar\'e dual to $T$. $e_0$ is a
(1,1)-form for suitable complex structure on $S_H$. We may now expand
the K\"ahler form on $S_H$ in terms of a basis of 2-forms on $S_H$ one
of which is $e_0$. That is
\begin{equation}
  J = \sum_i J_ie_i.
\end{equation}
We claim that the limit $\Phi_{\text{IIA}}\to-\infty$ is equivalent to
taking $J_0\to\infty$.

This is much simpler to state in the case that $S_H$ is an elliptic
fibration with a section, $\pi_H:S_H\to B$. Now taking
$\Phi_{\text{IIA}}\to-\infty$ is equivalent to taking the area of the
section (or base) to be very large.

This may be proven by using a fibre-wise duality argument much along
the lines of \cite{AL:ubiq}. In \cite{AL:ubiq} it was argued that
taking the heterotic string to be weakly-coupled is the same as taking
the section of $X$ as a K3-fibration to be very large. Here we are
arguing the converse --- the weakly-coupled type IIA string is dual to
the heterotic string on an elliptically-fibred K3 surface with a big
section. The fibre-wise duality argument of \cite{AL:ubiq} works
equally well in this case.

To proceed further we will assume that the heterotic K3 surface, $S_H$,
is an elliptic fibration with a section. Note that this will reduce
the number of moduli we are allowed to probe --- not all K3 surfaces
are elliptic with a section. Fortunately we will still be able to
reach a boundary where all quantum corrections disappear.

We have argued above that this section of $S_H$ is very large. If we
assume that this K3 is generic (i.e., all fibres are of Kodaira type
$\mathrm{I}_1$) then we may take the volume of $S_H$ to be very large
by making sure that the elliptic fibres have large area.

Again we may let fibre-wise duality suggest an interpretation of such
a limit in the type IIA picture. We follow an idea first explained in
\cite{FMW:F} following the work of \cite{MV:F2}. Consider a dual pair
of a heterotic string compactified on $T^4$ and a type IIA string
compactified on a K3 surface. The map between the moduli describing
these two compactifications is completely known \cite{me:lK3}. We may
thus ask what happens on the type IIA side if we take a 2-torus within
the 4-torus on the heterotic side to have very large area. The result
is that the K3 surface undergoes a ``stable degeneration''. Precisely
what the stable degeneration is depends upon whether we wish to
describe the $E_8\times E_8$ heterotic string or the $\spnh$ heterotic
string on a large $T^2$. The $\spnh$ case was described in
\cite{AM:po}. We will focus on the $E_8\times E_8$ case which was
described in \cite{FMW:F}.

The result is that the K3 surface becomes a union of two rational
elliptic surfaces intersecting along an elliptic curve. We will
describe the geometry of the rational elliptic surface 
in more detail later. In terms of an
elliptic fibration this union may be viewed as an elliptic fibration over
two $\P^1$'s touching at a point. See \cite{AM:po,me:MvF} for a more detailed
description of this.

Now let us return to the case of a \CY\ threefold, $X$, which is a
K3-fibration. The type IIA string on this space is dual to the
$E_8\times E_8$ heterotic string on $S_H\times E_H$. In order to take
the elliptic fibre of $S_H$ to large area the above argument suggests
that we should let each K3-fibre of $X$ undergo the corresponding
stable degeneration.

That is, $X$ becomes a fibration over $B$ where each fibre is now the
union of two rational elliptic surfaces joining along an elliptic
curve. 
We also want to view $X$ as an elliptic fibration over a
surface, $\Sigma$. In figure \ref{fig:deg} we depict our stable
degeneration in terms of this elliptic fibration. We need to introduce
quite a lot of notation to describe various aspects of this
degeneration.

\iffigs
\begin{figure}
  \centerline{\epsfxsize=14cm\epsfbox{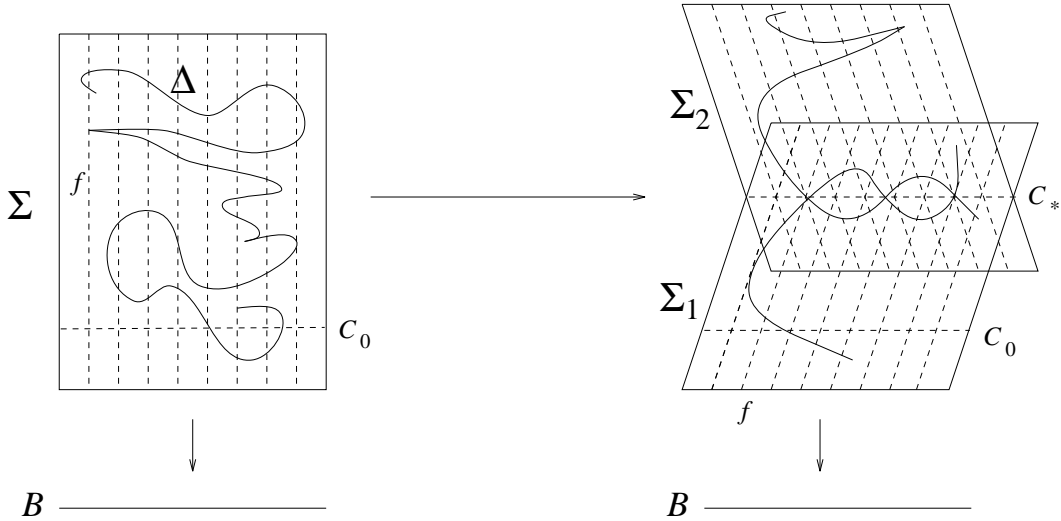}}
  \caption{The stable degeneration for $X$.}
  \label{fig:deg}
\end{figure}
\fi

Before the degeneration $\Sigma$ is a ruled surface. To be specific, $\Sigma$
is a $\P^1$-bundle over $\P^1$. We depict these $\P^1$-fibres by
vertical dotted lines in figure \ref{fig:deg}. We label a generic
vertical line by $f$. In the simplest case we may assume that $\Sigma$
is the Hirzebruch surface $\HS n$ along the lines of
\cite{MV:F,MV:F2}. Viewing $\HS n$ as a $\P^1$-fibration, we can
always find a section of self-intersection $-n$. We will denote a
generic such section by $C_0$ and we also denote it by a dotted line
in figure \ref{fig:deg}. Note that if $n>0$ then this section is
unique.

The elliptic fibration $\pi_F:X\to\Sigma$ is not smooth. Over a
discriminant locus, $\Delta\subset\Sigma$, the fibres degenerate. We
depict this locus by a solid line in figure \ref{fig:deg}.

Let $X^\sharp$ denote the degenerated version of $X$.
When we go to the stable degeneration, every $f$-line will break
into two lines intersecting at a point. Thus, the base of our elliptic
fibration becomes a bundle over $B$ with generic fibre given by {\em two\/}
lines. Restricting the elliptic fibration to either of these two lines
will give us a rational elliptic surface. Thus $X^\sharp$ is a
fibration over $B$ with fibre given by two rational elliptic surfaces
intersecting along an elliptic curve. Equally, $X^\sharp$ is the
union of two spaces, $X_1$ and $X_2$, intersecting along a surface
which is an elliptic fibration. We denote this surface $S_*$
(and thus $X^\sharp=X_1\cup_{S_*} X_2$).
Meanwhile $\Sigma$ has become the union of two ruled surfaces,
$\Sigma_1$ and $\Sigma_2$, intersecting along a $\P^1$. We denote this
$\P^1$ by $C_*$. $S_*$ is then the restriction of the elliptic
fibration, $\pi_F^\sharp:X^\sharp\to(\Sigma_1\cup_{C_*}\Sigma_2)$, to $C_*$.

One may show \cite{AM:po} that if $\Sigma$ was given by $\HS n$ then
each of $\Sigma_1$ and $\Sigma_2$ are isomorphic to $\HS n$ and $C_*$
is a section of both $\Sigma_1$ and $\Sigma_2$. With
respect to $\Sigma_1$ we may chose $C_0$ to be disjoint from the
section $C_*$. In this case $C_*$ becomes a section in the class of $C_0$
within $\Sigma_2$. 

The discriminant locus will divide itself between $\Sigma_1$ and
$\Sigma_2$. Note that if the discriminant within $\Sigma_1$ intersects
$C_*$ then the elliptic fibration of $S_*$ will contain a bad
fibre. For consistency the discriminant within $\Sigma_2$ must
intersect $C_*$ at the same point, and with the same degree.

\subsection{The cohomology of the degeneration} \label{ss:coh}

Now that we know the way that $X$ degenerates at the boundary of the
moduli space we wish to consider how we can address the question as to how
well the type IIA string (or F-theory) picture and the $E_8\times E_8$
heterotic picture agree.

To do this we will focus on the Ramond-Ramond fields in the type IIA
string which live in $H^{\text{odd}}(X,\R/\Z)$. Assuming $b_1(X)=0$
this reduces to $H^3(X,\R/\Z)$. Let
us assume that $H_3(X)$ is torsion free.\footnote{If 
$H_3(X)$ contains torsion, the
Ramond-Ramond moduli may include discrete degrees of freedom. This can
lead to interesting problems in duality which are not fully understood
at this time. These were discussed in \cite{FHSV:N=2,me:flower}.}
This means that the
Ramond-Ramond fields live in the ``intermediate Jacobian'' of $X$
given by $H^3(X,\R)/H^3(X,\Z)$. We would therefore like to see how
$H^3(X,\Z)$ behaves as we perform the degeneration $X\to
X^\sharp$.

Formally one may analyze this degeneration in terms of the
Clemens-Schmid exact sequence \cite{Mor:CS}. Roughly speaking the
following happens. Consider first a 3-cycle $K\in H_3(X_1)$ which does
not intersect $S_*$. Clearly this should contribute to
$H_3(X^\sharp)$. When counting these contributions the only worry
might be that a 3-cycle in $X_1$ may be homologically the same as a
3-cycle in $X_2$. Actually this cannot happen since such an
equivalence would imply the existence of an element of $H_3(S_*)$, but
$H_3(\text{K3})=0$. Thus $H_3(X_1)\oplus H_3(X_2)\subset
H_3(X^\sharp)$.

\iffigs
\begin{figure}
  \centerline{\epsfxsize=10cm\epsfbox{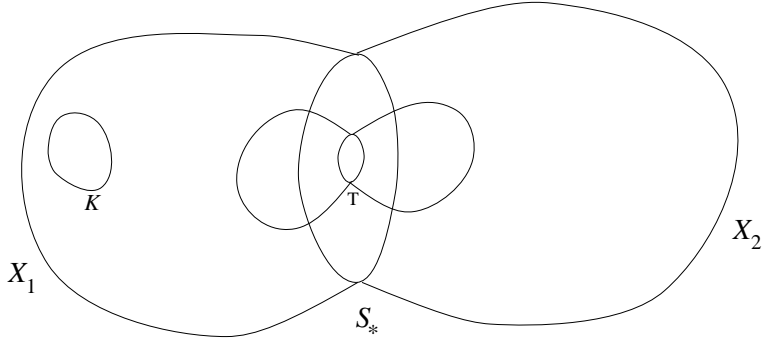}}
  \caption{The stable degeneration $X^\sharp$.}
  \label{fig:H3}
\end{figure}
\fi

The other contribution to $H_3(X^\sharp)$ arises in a more interesting
way. Suppose we have a 2-cycle $T$ in $H_2(S_*)$. There are natural
maps $f_1:H_2(S_*)\to H_2(X_1)$ and $f_2:H_2(S_*)\to H_2(X_2)$. Let us
suppose that $T$ lies in the kernel of both of these maps. This means
that $S_*$ is the boundary of a 3-chain that lives in $X_1$ and
another 3-chain that lives in $X_2$. Thus we may build an element of 
$H_3(X^\sharp)$ as shown in figure \ref{fig:H3}.

Let us assume that $S_*$ is in the form of an generic elliptic
fibration with a section. The elements of $H_2(S_*,\Z)$ generated by the
section and the generic fibre will not be in the kernel of $f_1$ and
$f_2$ as they map to the section and fibre of $X_1$ and $X_2$. The
other 20 two-cycles of $S_*$ will be in the kernel.
Let us denote this (self-dual) lattice of cycles by $M\cong
\Gamma_8(-1)\oplus\Gamma_8(-1)\oplus U\oplus U$, where $\Gamma_8(-1)$
is the root lattice of $E_8$, taken with a negative signature, and $U$
is the hyperbolic plane. $M$ can be thought of as the lattice of
cycles in $S_*$ which are not algebraic. It is known as the
``transcendental lattice''.
(Note that if $S_*$ is not generic 
then we may expect fewer than 20 cycles to contribute to the kernel
and the transcendental lattice will fall in rank.)
We have now argued that
\begin{equation}
  H^3(X^\sharp,\Z) \cong H^3(X_1,\Z)\oplus H^3(X_2,\Z) \oplus M.
\end{equation}

On the type IIA side therefore, our torus of Ramond-Ramond moduli has
factorized itself nicely into a product of three tori. What do these
three tori correspond to in the heterotic picture?

Firstly we have the heterotic K3 surface itself, $S_H$. We know the
complex structure of $S_H$ is given by the complex structure of
$S_*$. Since we have taken the volume of $S_H$ to be infinite it would
be unreasonable to assume we have anything left of the K\"ahler form data
of $S_H$. What remains then is the $B$-field. The obvious thing to do
is to associate $M$ with part of the moduli space of $B$-fields. This
is natural since $M$ was derived from $H^2(S_*,\Z)$. Note that we only
see the $B$-fields associated to the 20 transcendental cycles --- we
have ``lost'' two of the degrees of freedom in the $B$-field
associated to the section and fibre of $S_H$.

Next we have the two $E_8$-bundles. As we shall see in the next
section, and has been noted in \cite{FMW:F,CD:F4,Don:HetF}, the remaining
moduli of $X_1$ taken together with
the torus $H^3(X_1,\R)/H^3(X_1,\Z)$ produces the
hyperk\"ahler moduli space of one $E_8$ bundle while the corresponding
data on $X_2$ produces the other.

We have therefore achieved our goal. By going to this boundary in the
moduli space of theories we have been able to identify the moduli
space of theories exactly with the type IIA interpretation and exactly
with the heterotic interpretation. This has been at a price
of course. On the type IIA side we have lost some of the deformations
of $X$ and some of $H^3(X)$ --- resulting in a loss of Ramond-Ramond
moduli. On the heterotic string side we have lost the K\"ahler moduli
of $S_H$ and we have lost some deformations of complex structure by
demanding that $S_H$ be elliptic with a section. We also have only 20
of the original 22 $B$-field degrees of freedom. The boundary we have
reached has real codimension 24.


\section{The Bundle}    \label{s:bun}

\def\cG{\mathcal{G}}

\subsection{The \MW\ group}    \label{ss:MW}

In this section we will discuss the $E_8$-bundle moduli of the
heterotic string in the type IIA language. The ideas we present in
this section are essentially contained in
Friedman, Morgan, and Witten's work \cite{FMW:F} as well
as \cite{BJPS:F,CD:F4,Don:HetF}. We present here 
some explicit construction methods and we discuss
some examples to show how easily the duality map can be seen.
We also need to introduce an important lattice, $\Upsilon$, for
section \ref{s:disc}.
Our method will focus on the relationship between the associated
spectral curve of the bundle and the \MW\ group of rational elliptic surfaces.

$X_1$ is an elliptic fibration over $\Sigma_1\cong\HS
n$. Deformations of this threefold, together with its intermediate
Jacobian, will give 
the moduli space of one of the $E_8$-bundles. Again we denote by $C_0$
a section of $\HS n$ with self-intersection $-n$. $C_*$ is another
section, disjoint from $C_0$, with self-intersection $+n$. We denote a
generic $\P^1$-fibre of $\HS n$ by $f$. We thus have the
intersection relations $[C_0].[C_0]=-n$, $[C_0].[f]=1$, and $[f].[f]=0$.

Now write $X_1$ in Weierstrass form
\begin{equation}
  y^2 = x^3 + ax +b,	\label{eq:Wei}
\end{equation}
where $a$ and $b$ are sections of line bundles over $\Sigma_1$. Let
$A$ and $B$ denote the divisor classes corresponding to these
bundles. The discriminant is given by $4a^3 +
27b^2$. Let us denote the class of the discriminant by
$\Delta$. Sometimes we will be sloppy with our notation and refer to
the discriminant locus itself by $\Delta$.

One may show that \cite{AM:po}
\begin{equation}
\begin{split}
[C_*] &= [C_0] + n[f]\\
A &= 4[C_0] + 8[f]\\
B &= 6[C_0] + 12[f]\\
\Delta &= 12[C_0] + 24[f].
\end{split}  \label{eq:classes}
\end{equation}

In the generic case, $\Delta$ collides with $f$ transversely at 12
points. This means that, as expected, we obtain an rational elliptic
surface by restricting the elliptic fibration to $f$. In Kodaira
language this corresponds to building a rational elliptic surface by
an elliptic fibration with twelve $\mathrm{I}_1$ fibres.

There are many ways of building a rational elliptic surface in
Weierstrass form. Indeed, Persson has listed all 289 ways explicitly
\cite{Pers:RES}. Of central importance to us will be the notion of the
\MW\ group of this surface --- i.e., the group of sections
taking our given section as the identity. The \MW\ group of all the
possible rational elliptic surfaces was also determined in \cite{Pers:RES}.

One may show that the homology of 2-cycles of the rational elliptic surface
naturally decomposes into three parts:
\begin{enumerate}
\item The identity section and the fibre.
\item Components of fibres not touching the identity section.
\item Elements generated from the \MW\ group.
\end{enumerate}
The Picard lattice of the rational elliptic surface is isomorphic to
$\Gamma_8(-1)\oplus U$ of which the first part accounts for $U$. This,
the components of fibres not touching the 0-section and the
\MW\ group must account for $\Gamma_8(-1)$. Components of
fibres missing the 0-section arise when we have bad fibres in the
Kodaira classification other than $\mathrm{I}_1$ and II. These are
the two-cycles which are shrunk down to zero size to generate enhanced
gauge symmetries in the F-theory limit as in
\cite{MV:F}. Alternatively in type IIA language, these are the
two-cycles which are shrunk down to zero size in order to switch off
the Wilson lines around the heterotic 2-torus, $E_H$, again generating
an enhanced gauge symmetry. 

Since these 2-cycles which miss the 0-section are associated to
(perturbative) enhanced gauge symmetry, their existence must
correspond to the actual structure group of our supposed $E_8$-bundle
actually being less than we thought. To be precise, let the actual
structure group of the bundle be $\cG$. Then the observed gauge symmetry
is given by the centralizer of $\cG\subset E_8$. 
Note that the \MW\ contribution towards $H_2$ is the
complement of this enhanced gauge symmetry contribution within
$\Gamma_8(-1)$. This tells us that the
\MW\ group part of the contribution must be closely associated
to $\cG$. Indeed we shall now see how the \MW\
group of each rational elliptic surface is key in constructing the
heterotic bundles.

In the generic case we have twelve $\mathrm{I}_1$ fibres and no
enhanced gauge symmetry. In this case the \MW\ group is rank 8
and the structure group really is $\cG\cong E_8$. As this is the
hardest to visualize 
let us try something a little more manageable. 

At the other extreme of the bundle with a structure group given by the
complete $E_8$, we may try to build a bundle with a trivial structure
group. This may be done and corresponds to ``point-like
instantons''. This case is somewhat subtle and we postpone it for a
while.

\subsection{SU(2)-bundles}  \label{ss:A1}

Instead we try to build an SU(2)-bundle. Since $\mf{e}_7\oplus
\mf{sl}(2)$ is a maximal subalgebra of $\mf{e}_8$, the centralizer of
SU(2)$\subset E_8$ is $E_7$.\footnote{Note that $E_7\times$SU(2) is
{\em not\/} a subgroup of $E_8$ however!} Thus if the heterotic string
is compactified on an SU(2)-bundle, there will be an unbroken $E_7$
gauge symmetry. The rational elliptic curve we desire as our generic
fibre over a point in $B$ is given by

\begin{equation}
\setlength{\unitlength}{0.006250in}%
\begin{picture}(315,76)(180,560)
\thinlines
\put(260,600){\makebox(0,0){$\times$}}
\put(420,580){\makebox(0,0){$\times$}}
\put(460,600){\makebox(0,0){$\times$}}
\put(380,573){\makebox(0,0){$\times$}}
\put(180,610){
}
\put(260,615){\makebox(0,0)[lb]{$\mathrm{III}^*$}}
\put(420,548){\makebox(0,0)[lb]{$\mathrm{I}_1$}}
\put(460,568){\makebox(0,0)[lb]{$\mathrm{I}_1$}}
\put(380,541){\makebox(0,0)[lb]{$\mathrm{I}_1$}}
\put(155,600){\makebox(0,0)[lb]{$f$}}
\end{picture}
\end{equation}

This has Mordell-Weil group isomorphic to $\Z$, i.e., of rank 1. This
means that there is another section, $\sigma_1$, of this rational
elliptic surface which generates the \MW\ group. The
difference between this section and our zero section, $\sigma_0$,
generates part of $H_2$ of the surface (see, for example,
\cite{MirPer:ell}), $[\sigma_1-\sigma_0]\in H_2$. 

Remember that we are building $X_1$ as a rational elliptic surface
fibration over $B$. There can be monodromy in this fibration such that
our 2-cycle $[\sigma_1-\sigma_0]$ is not an invariant homology
cycle. It is possible that parallel transport along a closed loop in
$B$ may map $[\sigma_1-\sigma_0]$ to
$-[\sigma_1-\sigma_0]=[\sigma_{-1}-\sigma_0]$, where 
$\sigma_{-1}$ is another section in the \MW\ group. Under the
group law, $\sigma_{-1}$ is the inverse of $\sigma_1$.

Now this rational elliptic surface intersects $S_*$ along an elliptic
curve. This elliptic curve is the elliptic fibre over the point in
$\Sigma_1$ given by the collision of $C_*$ with the $f$-line we have
taken to build our rational elliptic surface. The sections $\sigma_1$
and $\sigma_{-1}$ will each intersect this elliptic curve at two
points, $P_1$ and $P_{-1}$. Consider now the whole family of rational
elliptic surfaces 
over $B$. Transporting $P_1$ and $P_{-1}$ around the whole of $B$ will
build a double cover $\pi_s:C\to B$. This curve, $C$, is the {\em
spectral curve}. 

\begin{figure}
\begin{center}
\setlength{\unitlength}{0.008750in}%
\begin{picture}(365,210)(140,575)
\thinlines
\put(140,600){\line( 1, 0){360}}
\multiput(330,780)(0.00000,-8.03922){26}{\line( 0,-1){  4.020}}
\multiput(230,580)(0.00000,8.03922){26}{\line( 0, 1){  4.020}}
\multiput(420,780)(0.00000,-7.84314){26}{\line( 0,-1){  3.922}}
\multiput(160,740)(8.00000,0.00000){43}{\line( 1, 0){  4.000}}
\multiput(165,580)(0.00000,8.03922){26}{\line( 0, 1){  4.020}}
\put(140,580){
}
\put(165,680){
}
\put(425,755){\makebox(0,0)[lb]{$f_2$}}
\put(505,590){\makebox(0,0)[lb]{$C_0$}}
\put(505,735){\makebox(0,0)[lb]{$C_*$}}
\put(450,605){\makebox(0,0)[lb]{III$^*$}}
\put(235,755){\makebox(0,0)[lb]{$f_1$}}
\put(335,755){\makebox(0,0)[lb]{$f$}}
\put(170,755){\makebox(0,0)[lb]{$f_3$}}
\put(365,640){\makebox(0,0)[lb]{I$_1$}}
\end{picture}
\end{center}
  \caption{The discriminant for an SU(2)-bundle.}
\label{fig:A1}
\end{figure}
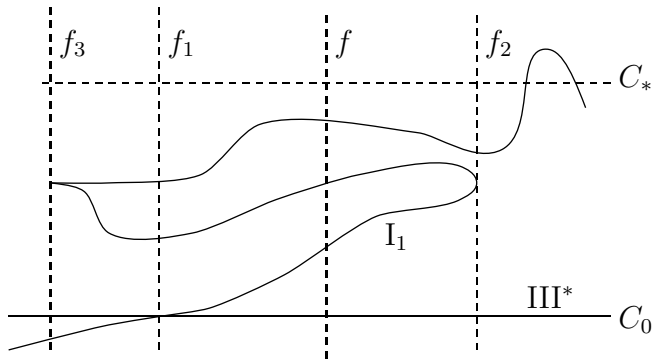

Globally, in terms of $\Sigma_1$, our picture for this elliptic
fibration looks like figure 
\ref{fig:A1}. Over the generic $f$-line, labelled by $f$ in the
figure, we have one III$^*$ fibre and three I$_1$ fibres as desired
for our generic rational elliptic surface. In the generic case there
are three things that 
can happen to spoil this for some particular $f$-lines. 
In the following recall that $n$ refers to the geometry of
$\Sigma_1\cong\HS n$.
\begin{enumerate}[(a)]
\item The curve of I$_1$'s may collide with the line $C_0$ of III$^*$
fibres. This does not occur transversely --- each collision has
multiplicity 3. One may also show from intersection theory that such a
collision occurs $8-n$ times \cite{MV:F2}. When the elliptic fibration
is restricted to an $f$-line through this collision (labelled $f_1$ in
the figure) we obtain a rational elliptic surface with one II$^*$
fibre and two I$_1$ fibres.
\item The curve of I$_1$'s may be tangential to the $f$ direction. 
When the elliptic fibration is restricted to such an $f$-line
(labelled $f_2$ in the figure) we obtain a rational elliptic surface
with one III$^*$ fibre, one I$_2$ fibre and one I$_1$ fibre.
\item The curve of I$_1$'s may have a cusp. An $f$-line passing
through such a point is labelled by $f_3$ in the diagram. Now the
resulting rational elliptic surface has one III$^*$ fibre, one
II fibre and one I$_1$ fibre. 
\end{enumerate}

In the case (a), the \MW\ group becomes trivial. This must
mean that the sections $\sigma_1$ and $\sigma_{-1}$ must coincide with
$\sigma_0$ at this point over $B$. In case (b), the \MW\ group
becomes $\Z_2$. For this surface both $\sigma_1$ and $\sigma_{-1}$ pass
through a 2-torsion point in the elliptic fibre over $C_*$. In both of these
cases, $\sigma_1$ and $\sigma_{-1}$ coincide and we have a branch
point of $\pi_s:C\to B$. In case (c), the \MW\ group
remains equal to $\Z$ and $\pi_s:C\to B$ is not branched.

We may calculate the genus of $C$ from the number of branch
points. As mentioned earlier, the are $8-n$ collisions from case
(a). 
To count case (c) we look for collisions between $A$ and $B$ given by
(\ref{eq:classes}). For a similar computation see the computation in
section 6.6 of \cite{me:lK3}. The result is $20-n$.

To calculate (b) we need to write our discriminant more
explicitly. 
Let $s$ and $t$ be affine coordinates on $\Sigma_1\cong\HS n$. We let
$s$ be a coordinate in the ``fibre'' direction and $t$ be in the ``base''
direction. 
The curve of I$_1$ fibres in figure \ref{fig:A1} is in the class
$\Delta - 9[C_0]=3[C_0]+24[f]$. This means that the polynomial,
$\delta_1$, whose zero gives the curve of I$_1$ fibres can be written
in the form
\begin{equation}
  \delta_1 = s^3f_{24}(t) + s^2f_{24-n}(t) + sf_{24-2n}(t) + f_{24-3n}(t),
		\label{eq:cubA1}
\end{equation}
where $f_m(t)$ represents some generic polynomial of degree $m$ in
$t$. Similar equations appeared in \cite{MV:F2}. The discriminant of
$\delta_1$ with respect to $s$ is a polynomial of degree $96-6n$ in
$t$. This measures the number of points over the base where the three
solutions of the cubic equation given by (\ref{eq:cubA1})
coincide. This occurs for cases (b) and (c). 

It is easy to show that a generic cusp of type (c) will contribute 3
towards the above discriminant while the tangency of type (b) will
contribute only 1. Thus the number of occurrences of case (b) is
$96-6n-3(20-n)=36-3n$. The number of branch points is therefore given
by $(8-n) + (36-3n)=44-4n$ from which is it follows that $C$ has genus
$\ff12(44-4n)-1=21-2n$.

We wish to claim that there is a natural identification of the
Jacobian of $C$, $H^1(C,\R/\Z)$, with the moduli space of
Ramond-Ramond fields $H^3(X_1,\R/\Z)$. This is easy given our
construction of the spectral curve. First let us ask how we see
$H_3(X_1)$ in terms of the fibration $p_1:X_1\to B$. Such 3-cycles are
built from transporting 2-cycles within fibre around non-contractable
1-cycles in $B$. These 2-cycles must be non-monodromy-invariant to
build a homologically nontrivial element of $H_3(X_1)$. These 2-cycles
come exactly from the \MW\ group of the fibre of $p_1$. We may
now ask how we build elements of $H_1(C)$ in terms of the fibration
$\pi_s:C\to B$. The answer is very similar except now we transport the
points in the fibre around loops in $B$. By construction however we
identify points in the fibre of $\pi_s$ with the 2-cycles coming from
the \MW\ group in the fibre of $p_1$. Thus $H_1(S)\cong
H_3(X_1)$ and so $H^1(C,\R/\Z)\cong H^3(X_1,\R/\Z)$.

Actually we may do a little better than this. The Jacobian of a curve
is not only a torus but an {\em abelian variety\/}. That is, it is a
torus which admits a complex structure and can be embedded in some
complex projective space. One can argue that the same is true for the
intermediate Jacobian for $X_1$. One can then show that the Jacobian
of $C$, $\Jac(C)$, is isomorphic to the intermediate Jacobian of
$X_1$ as an algebraic variety (for example, see the Abel-Jacobi map of
\cite{Ty:Jac}). 

We may also phrase our construction more formally by using the
Leray spectral sequence of a fibration and noting that
$H^1(B,\pi_{s*}\Z)\cong H^1(B,R^2p_{1*}\Z)$.

We are done once we note that an SU(2)-bundle on $S_H$ is specified
uniquely by a given spectral curve, $C\subset S_H$, and a line bundle
(of a particular degree) $L\to C$. This was shown in \cite{FMW:F}. The
deformations of the spectral curve are given by the deformations of
$X_1$ and the moduli space of the line bundle $L$ are given by the
Jacobian of $C$ and thus the Ramond-Ramond fields.

As a final check let us compute the dimensions of the moduli
space. The moduli space of SU(2)-bundles on a K3 surface is a
hyperk\"ahler space whose quaternionic dimension is given by
Hirzebruch-Riemann-Roch as $2c_2-3$. 
Similarly the moduli space of $L$ together with its Jacobian forms a
hyperk\"ahler space as in Hitchin's construction \cite{Hit:bun}. The
quaternionic dimension of this moduli space is given by the genus of
$L$ which is $21-2n$. Thus we are in agreement if $c_2=12-n$ as
expected \cite{MV:F}.

\subsection{$G_2$-bundles}  \label{ss:G2}

Having explicitly mapped the moduli space of SU(2)-bundles on $S_H$ to
its type IIA dual picture in terms of $X_1$, we may try to do the same
for a larger structure group, $\cG$. The obvious thing to do is to
slightly relax our constraint above that we have an unbroken $E_7$
gauge symmetry. 

As the next case we use the generic rational elliptic surface
in the fibre of $p_1:X_1\to B$ of the following form:
\begin{equation}
\setlength{\unitlength}{0.006250in}%
\begin{picture}(315,76)(180,560)
\thinlines
\put(260,600){\makebox(0,0){$\times$}}
\put(420,580){\makebox(0,0){$\times$}}
\put(460,600){\makebox(0,0){$\times$}}
\put(380,573){\makebox(0,0){$\times$}}
\put(340,573){\makebox(0,0){$\times$}}
\put(180,610){
}
\put(260,615){\makebox(0,0)[lb]{$\mathrm{IV}^*$}}
\put(420,548){\makebox(0,0)[lb]{$\mathrm{I}_1$}}
\put(460,568){\makebox(0,0)[lb]{$\mathrm{I}_1$}}
\put(380,541){\makebox(0,0)[lb]{$\mathrm{I}_1$}}
\put(340,541){\makebox(0,0)[lb]{$\mathrm{I}_1$}}
\put(155,600){\makebox(0,0)[lb]{$f$}}
\end{picture}
\end{equation}

The type IV$^*$ fibre is associated to an $E_6$ singularity and so one
might at first suppose that we are going to see an unbroken $E_6$
gauge symmetry. This is not the case however thanks to monodromy
\cite{AG:sp32}. In fact there is more than one way in which this case
will be quite a bit more subtle than the SU(2) case above.

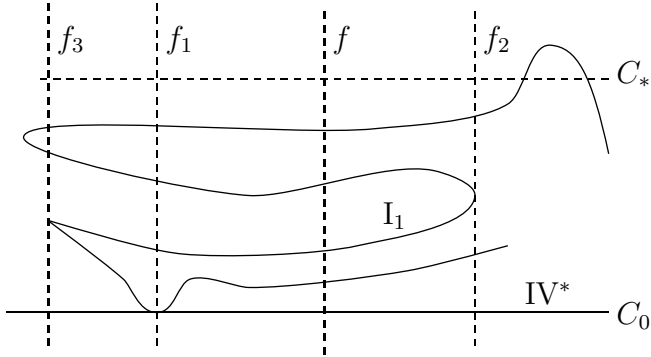
\begin{figure}
\begin{center}
\setlength{\unitlength}{0.008750in}%
\begin{picture}(365,210)(140,575)
\thinlines
\put(140,600){\line( 1, 0){360}}
\multiput(330,780)(0.00000,-8.03922){26}{\line( 0,-1){  4.020}}
\multiput(230,580)(0.00000,8.03922){26}{\line( 0, 1){  4.020}}
\multiput(420,780)(0.00000,-7.84314){26}{\line( 0,-1){  3.922}}
\multiput(160,740)(8.00000,0.00000){43}{\line( 1, 0){  4.000}}
\multiput(165,580)(0.00000,8.03922){26}{\line( 0, 1){  4.020}}
\put(440,640){
}
\put(165,655){
}
\put(425,755){\makebox(0,0)[lb]{$f_2$}}
\put(505,590){\makebox(0,0)[lb]{$C_0$}}
\put(505,735){\makebox(0,0)[lb]{$C_*$}}
\put(235,755){\makebox(0,0)[lb]{$f_1$}}
\put(335,755){\makebox(0,0)[lb]{$f$}}
\put(170,755){\makebox(0,0)[lb]{$f_3$}}
\put(450,605){\makebox(0,0)[lb]{IV$^*$}}
\put(365,650){\makebox(0,0)[lb]{I$_1$}}
\end{picture}
\end{center}
  \caption{The discriminant for a $G_2$-bundle.}
\label{fig:G2}
\end{figure}

In figure \ref{fig:G2} we draw the discriminant locus of the elliptic
fibration $X_1\to\Sigma_1$ again. Now we have the following collisions:
\begin{enumerate}[(a)]
\item The curve of I$_1$'s may collide with the line $C_0$ of IV$^*$
fibres. This does not occur transversely --- each collision has
multiplicity 2. One may also show from intersection theory that such a
collision occurs $2(6-n)$ times. When the elliptic fibration
is restricted to an $f$-line through this collision (labelled $f_1$ in
the figure) we obtain a rational elliptic surface with one III$^*$
fibre and three I$_1$ fibres.
\item The curve of I$_1$'s may be tangential to the $f$ direction. 
When the elliptic fibration is restricted to such an $f$-line
(labelled $f_2$ in the figure) we obtain a rational elliptic surface
with one IV$^*$ fibre, one I$_2$ fibre and two I$_1$ fibres.
\item The curve of I$_1$'s may have a cusp. An $f$-line passing
through such a point is labelled by $f_3$ in the diagram. Now the
resulting rational elliptic surface has one IV$^*$ fibre, one
II fibre and two I$_1$ fibres. 
\end{enumerate}

Over a generic $f$, the rational elliptic surface has a \MW\
group equal to $\Z\oplus\Z$. Over $f_1$ and $f_2$ in cases (a) and (b) the
\MW\ group becomes $\Z$. Thus we again have some kind of
branched curve over $B$ with branch points at the locations of $f_1$
and $f_2$. 

What kind of monodromy will we get for these two-cycles coming from
the \MW\ group? We may address this question from our
knowledge that the total lattice of cycles generated by the
\MW\ group and the reducible fibres is $\Gamma_8(-1)$ as
explained earlier. 
Over the generic curve $f$ in figure \ref{fig:G2}
we have the type IV$^*$ fibre which contributes an $\mf{e}_6$ root lattice
of cycles to the Picard lattice. Thus the \MW\ group must
contribute the orthogonal complement of this $\mf{e}_6$ root lattice which
is the $\mf{sl}(3)$ root lattice. The monodromy around the $f_2$ fibre
in figure \ref{fig:G2} leaves the $\mf{e}_6$ root lattice alone and so
we may obtain an element of the Weyl group of $\mf{sl}(3)$ acting on
the \MW\ part.

The monodromy around the $f_1$ line is more interesting. It is known
that the monodromy around this line is not trivial on the IV$^*$ fibre
and thus not trivial on the $\mf{e}_6$ root lattice. 
Actually the monodromy acts as an outer automorphism of $\mf{e}_6$.
Since the
full $\Gamma_8(-1)$ lattice of cycles must be mapped to itself by this
monodromy we need to know how such an outer automorphism acts within
the context of the $\mf{e}_8$ root lattice.

Let us label the simple roots of $\mf{e}_8$ as follows:
\begin{equation}
\setlength{\unitlength}{0.006875in}%
\begin{picture}(375,108)(110,690)
\thinlines
\put(120,720){\circle{10}}
\put(180,720){\circle{10}}
\put(240,720){\circle{10}}
\put(300,720){\circle{10}}
\put(360,720){\circle{10}}
\put(420,720){\circle{10}}
\put(480,720){\circle{10}}
\put(240,780){\circle{10}}
\put(125,720){\line( 1, 0){ 50}}
\put(185,720){\line( 1, 0){ 50}}
\put(245,720){\line( 1, 0){ 50}}
\put(365,720){\line( 1, 0){ 50}}
\put(425,720){\line( 1, 0){ 50}}
\put(240,775){\line( 0,-1){ 50}}
\put(305,720){\line( 1, 0){ 50}}
\put(110,690){\makebox(0,0)[lb]{$\alpha_1$}}
\put(250,770){\makebox(0,0)[lb]{$\alpha_2$}}
\put(170,690){\makebox(0,0)[lb]{$\alpha_3$}}
\put(230,690){\makebox(0,0)[lb]{$\alpha_4$}}
\put(290,690){\makebox(0,0)[lb]{$\alpha_5$}}
\put(350,690){\makebox(0,0)[lb]{$\alpha_6$}}
\put(410,690){\makebox(0,0)[lb]{$\alpha_7$}}
\put(470,690){\makebox(0,0)[lb]{$\alpha_8$}}
\end{picture}
\end{equation}
Now introduce some more roots
\begin{equation}
\begin{split}
\alpha_9 &= -2\alpha_1-3\alpha_2-4\alpha_3-6\alpha_4-5\alpha_5
	-4\alpha_6-3\alpha_7-2\alpha_8\\
\alpha_{10} &= -\alpha_8-\alpha_9\\
\alpha_{11} &= -\alpha_1-2\alpha_2-2\alpha_3-3\alpha_4-2\alpha_5-\alpha_6.
\end{split}
\end{equation}
We may draw a Dynkin-like diagram representing the angles between some
of these roots as follows:
\begin{equation}
\setlength{\unitlength}{0.00050000in}%
\begin{picture}(4200,2925)(1951,-2602)
\thinlines
\put(2101,-961){\circle{150}}
\put(3001,-961){\circle{150}}
\put(3601,-361){\circle{150}}
\put(3601,-1561){\circle{150}}
\put(4201,239){\circle{150}}
\put(4201,-2161){\circle{150}}
\put(5401,-961){\circle{150}}
\put(6001,-361){\circle{150}}
\put(6001,-1561){\circle{150}}
\put(2176,-961){\line( 1, 0){750}}
\put(3061,-901){\line( 1, 1){495}}
\put(3661,-301){\line( 1, 1){495}}
\put(3541,-1501){\line(-1, 1){495}}
\put(4141,-2101){\line(-1, 1){495}}
\put(5461,-901){\line( 1, 1){495}}
\put(6001,-436){\line( 0,-1){1050}}
\put(4261,179){\line( 1,-1){1080}}
\put(4261,-2116){
}
\put(5461,-1021){
}
\put(1951,-1336){\makebox(0,0)[lb]{\smash{$\alpha_{3}$}}}
\put(2881,-1336){\makebox(0,0)[lb]{\smash{$\alpha_{4}$}}}
\put(3691,-601){\makebox(0,0)[lb]{\smash{$\alpha_{5}$}}}
\put(4381,119){\makebox(0,0)[lb]{\smash{$\alpha_{6}$}}}
\put(4141,-2551){\makebox(0,0)[lb]{\smash{$\alpha_{11}$}}}
\put(5281,-1366){\makebox(0,0)[lb]{\smash{$\alpha_{7}$}}}
\put(6151,-451){\makebox(0,0)[lb]{\smash{$\alpha_{8}$}}}
\put(6151,-1681){\makebox(0,0)[lb]{\smash{$\alpha_{10}$}}}
\put(3511,-1981){\makebox(0,0)[lb]{\smash{$\alpha_{2}$}}}
\end{picture}
\label{eq:G2F4}
\end{equation}
Here a dotted line indicates that the inner product is negative that of a
solid line. Clearly this diagram shows that we have a symmetry of the
root system by reflecting (\ref{eq:G2F4}) about a horizontal line and
taking $\alpha_7\to-\alpha_7$.

This symmetry of the root system is an element of the Weyl group of
$\mf{e}_8$. The subdiagram to the left of $\alpha_7$ in
(\ref{eq:G2F4}) is the Dynkin diagram of $\mf{e}_6$ and this symmetry
corresponds to the outer automorphism. The subdiagram to the right of
$\alpha_7$ in (\ref{eq:G2F4}) is the Dynkin diagram of
$\mf{sl}(3)$. It must be then that the monodromy around $f_1$
exchanges the two simple roots of $\mf{sl}(3)$.

This is not in the Weyl group of $\mf{sl}(3)$. It is in the Weyl group
of $\mf{g}_2$ however. Indeed, if $W(\mf{h})$ is the Weyl group of
the Lie algebra $\mf{h}$ then
\begin{equation}
  0\to W(\mf{sl}(3)) \to W(\mf{g}_2) \to \Z_2\to0,
\end{equation}
where this latter $\Z_2$ is precisely the exchange generated above by the
monodromy. 

From the heterotic language we can argue that the bundle we are
constructing here is indeed a $G_2$-bundle and not an
SU(3)-bundle. This is because the monodromy causes the effective gauge
group to be observed to be $F_4$ rather than $E_6$
\cite{AG:sp32}. Since there is a maximal subgroup
\begin{equation}
  E_8 \supset G_2\times F_4,   \label{eq:E8G2}
\end{equation}
we should expect an $F_4$ gauge symmetry to be preserved by a
$G_2$-bundle.
Indeed the way that one can demonstrate the maximal embedding of
(\ref{eq:E8G2}) is via an argument along the lines of the diagram in 
(\ref{eq:G2F4}).

This example shows a rule (which follows from Donagi's construction
\cite{Don:spec}) that the
Weyl group of the structure group, $\cG$, of the bundle is generated by the
monodromy on the \MW\ part of the Picard lattice of the
rational elliptic surfaces.

We also see that if the $2(6-n)$ collisions of the type along $f_1$ in
figure \ref{fig:G2} coalesce into pairs then we lose any monodromy of
this type and reduce the monodromy group to that of SU(3). This is in
agreement with the similar statement that such a pairing results in an
$E_6$ unbroken gauge symmetry.

Take one of the generators of the Mordell-Weil group, $\Z\oplus\Z$, of the
rational elliptic surface in a generic fibre. This has an orbit of six
elements via the monodromy around $B$. We may therefore build a
six-fold branched cover $\pi_s:C\to B$. This again we may call the
``spectral curve'' as in section \ref{ss:A1}. It is important to note
however that this spectral cover for the $G_2$ case is not quite as
simple as that for the SU(2) case. 

It is now untrue that $H^1(B,\pi_{s*}\Z)\cong H^1(B,R^2p_{1*}\Z)$ and
so we cannot identify the Ramond-Ramond moduli of the type IIA string
with the Jacobian of the spectral curve. This story is familiar one in
the theory of branched covers and has been studied extensively by
Donagi in the current context \cite{Don:spec}.

The general idea is that one may identify a ``Prym'' which is a
subspace of the Jacobian of the spectral curve. This Prym is also an
abelian variety. One can then show that
this Prym matches the Ramond-Ramond moduli in terms of the type IIA
theory and that it correctly reproduces the moduli space of the bundle
on the heterotic side. The former was shown by Kanev \cite{Kan:delP}
and the latter by Donagi \cite{Don:F}.

The basic construction of the Prym is as follows. Let $W$ be the Weyl
group of $\cG$. The
spectral curve may be thought of as an $R$-cover of
$B$ where $R$ is some representation of $W$. That is to say, the
point-like fibres of the map $\pi_s:C\to B$ 
form a representation of $W$ under monodromy. The problem with $G_2$
that we did not have with $\SU(2)$ in the last section, is that to find
the agreement $H^1(B,\pi_{s*}\Z)\cong H^1(B,R^2p_{1*}\Z)$ we have the
{\em wrong\/} representation. In order to find agreement, we need the
natural representation of the Weyl group on the root space --- i.e.,
the representation with dimension equal to $\rank(\cG)$. 

Thus we replace one cover of $B$ with another one and try to take the
latter's Jacobian. We cannot do this literally as the root space
representation of the Weyl group is not a permutation representation.
Thus the supposed spectral curve for this representation cannot
actually be thought of as a branched cover.  Instead consider the
``cameral cover''. This is a ($W$-Galois) cover where the
representation is given by Weyl chambers. This is a $d$-fold cover of
$B$ where $d$ is the number of elements in $G$. Let $\widetilde C$ be
the cameral cover. Let $\Lambda$ be the root lattice of $\cG$ and let
$\Jac(\widetilde C)$ be the Jacobian of $\widetilde C$. Clearly $W$
has an action on both $\Lambda$ and $\Jac(\widetilde C)$. The Prym we
desire may be written as
\begin{equation}
  P = \Hom_W(\Lambda,\Jac(\widetilde C)),
\end{equation}
where the $W$ subscript means that $p(g\lambda)=gp(\lambda)$ for all
$p\in P$, $g\in W$, $\lambda\in\Lambda$. 
In general this is a disconnected set of abelian varieties. It is an
interesting question how to deal with the cases where there is more
than one component in this Prym. Here we will assume we are dealing
with the component connected to the identity if there is more than one
component.
In other words there is an
$l$ such that
\begin{equation}
  P = \C^l/\Upsilon,
\end{equation}
where $\Upsilon$ is a lattice of dimension $2l$. A little algebra
shows that
\begin{equation}
  \Upsilon = \Hom_W(\Lambda,H^1(\widetilde C,\Z)).
\end{equation}
 In the simplest case of $\cG\cong\SU(N)$, 
the Weyl group is the symmetric group on $N$ objects, the spectral
curve is an $N$-fold cover of $B$,
the Prym is equal to the
Jacobian of the spectral cover%
\footnote{Actually we should be more careful and say there exists a
spectral curve such that the Jacobian of this curve is isomorphic to
the Prym. The construction from the \MW\ group produces the spectral curve
from the permutation action of $W$ on the roots of $\cG$ rather than
the $N$-dimensional permutation representation. These two spectral
curves happened to coincide for $\cG\cong\SU(2)$.}
and $\Upsilon=H^1(C,\Z)$. For general
$\cG$ life is more complicated.

If the situation is sufficiently generic that each monodromy
corresponds to a Weyl reflection in one root then $l$ may be
determined. Let there be $b$ branch points. We show in the appendix that
\begin{equation}
  l = \ff12 b - \rank(\cG).
\end{equation}

Applying this formula to our $G_2$-bundle, we have $2(6-n)$ collisions
of type (a) and it is straight-forward to show that there are $60-6n$
collisions of type (b). Thus $b=72-8n$ and $l=34-4n$. This value of
$l$ agrees with 
the quaternionic dimension of the moduli space of $G_2$-bundles on a K3
with $c_2=12-n$ again.

It is a simple matter now to extend this construction to
$\cG=\SU(3)$. If we allow the collisions of type (a) to pair up, we
lose the monodromy which elevated $\SU(3)$ to $G_2$. Thus we should
only have an $\SU(3)$-bundle. Now we only count the type (b)
collisions as branch points and so $b=60-6n$ and $l=28-3n$ again in
agreement with the direct bundle computation.

\subsection{Bundles with no structure group}  \label{ss:A0}

Now we want to try to build a bundle with no structure group at all. This
suggests picking the generic rational elliptic surface in $X_1$ to have a
trivial \MW\ group. This forces the rational elliptic surface
to be of the form
\begin{equation}
\setlength{\unitlength}{0.006250in}%
\begin{picture}(315,76)(180,560)
\thinlines
\put(260,600){\makebox(0,0){$\times$}}
\put(420,580){\makebox(0,0){$\times$}}
\put(460,600){\makebox(0,0){$\times$}}
\put(180,610){
}
\put(260,615){\makebox(0,0)[lb]{$\mathrm{II}^*$}}
\put(420,550){\makebox(0,0)[lb]{$\mathrm{I}_1$}}
\put(460,570){\makebox(0,0)[lb]{$\mathrm{I}_1$}}
\put(155,600){\makebox(0,0)[lb]{$f$}}
\end{picture}
\end{equation}
or by letting the two $\mathrm{I}_1$'s coalesce to form a type II
fibre.

If every $f$ is of this form within $\Sigma_1$ then we will have a
whole curve of type $\mathrm{II}^*$ fibres. By the usual F-theory
recipe this means we have an unbroken $E_8$ gauge symmetry. This is
most reasonable if our bundle has no structure group! Of course, what
we are talking about here are the ``point-like instantons'' of
\cite{W:small-i,MV:F}.

\begin{figure}
\begin{center}
\setlength{\unitlength}{0.00041700in}%
\begin{picture}(10887,3543)(889,-3673)
\thinlines
\put(901,-3061){\line( 1, 0){4500}}
\put(3601,-661){
}
\put(5701,-1711){\vector( 1, 0){1200}}
\put(1501,-961){
}
\put(7501,-3286){\line( 1, 0){4125}}
\put(9601,-811){
}
\put(7651,-436){
}
\put(9226,-1336){
}
\put(1351,-2611){
}
\put(1576,-3211){
}
\put(7801,-2161){
}
\put(7876,-2836){
}
\put(3676,-811){\makebox(0,0)[lb]{\smash{$f$}}}
\put(5476,-3211){\makebox(0,0)[lb]{\smash{$C_0$}}}
\put(4576,-2611){\makebox(0,0)[lb]{\smash{I$_1$}}}
\put(4951,-3586){\makebox(0,0)[lb]{\smash{I$_1$}}}
\put(4951,-2986){\makebox(0,0)[lb]{\smash{II$^*$}}}
\put(5401,-1111){\makebox(0,0)[lb]{\smash{$C_*$}}}
\put(11701,-1936){\makebox(0,0)[lb]{\smash{I$_1$}}}
\put(11551,-2686){\makebox(0,0)[lb]{\smash{I$_1$}}}
\put(8776,-3211){\makebox(0,0)[lb]{\smash{II$^*$}}}
\put(11776,-3436){\makebox(0,0)[lb]{\smash{$C_0$}}}
\put(10726,-286){\makebox(0,0)[lb]{\smash{$f'$}}}
\put(11401,-661){\makebox(0,0)[lb]{\smash{$C_*$}}}
\end{picture}
\end{center}
  \caption{The blow-up for a generic point-like $E_8$ instanton.}
\label{fig:sm}
\end{figure}

The geometry of $E_8\times E_8$ point-like instantons in F-theory or
type IIA language has been explained in great length in various
places (for example, \cite{MV:F2,me:lK3}). The key point is that
within $\Sigma_1$, there are collisions between the curve of
$\mathrm{II}^*$ fibres and the curve of remaining $\mathrm{I}_1$
fibres within $\Delta$. A simple calculation in intersection theory
shows this happens at $12-n$ points. The resulting threefold, $X_1$,
produced by this fibration is highly singular over each of these
$12-n$ points. It may be smoothed out by blowing up each such
collision in $\Sigma_1$. Each such collision and the resulting blow-up
in shown in figure \ref{fig:sm}. In F-theory language these blow-ups give
tensor multiplets. Each of the $12-n$ collisions is associated to a
point-like instanton for the heterotic string on a K3 surface where
each instanton has $c_2=1$. Thus we have a total $c_2$ given by $12-n$
yet again.

What we would like to consider now is whether we actually
have the correct moduli space for these $12-n$ points. Each point-like
instanton should presumably have as its moduli space the K3 surface,
$S_H$, itself. We may restrict attention to such a single instanton by
fixing $n=11$.

To achieve the configuration of the discriminant locus in figure
\ref{fig:sm} we are required to restrict the forms of the polynomials
$a$ and $b$ in equation (\ref{eq:Wei}). A type $\mathrm{II}^*$ fibre
requires $a$ to vanish order at least 4 and $b$ to vanish order 5. 
The divisors class $B$ therefore divides into two parts $B=B'+5[C_0]$,
where $B'=[C_0]+12[f]$. The collisions in figure \ref{fig:sm} occur at
collisions between $B'$ and $C_0$. We are free to move these collisions
around by varying $b$.

It was argued in \cite{AM:po} that the location of these collisions
corresponded to the point-like instantons. To be more precise,
consider the elliptic fibration of $S_H\cong S_*$ given by
$\pi_H:S_H\to B$. We also have the fibration $p:\Sigma_1\to B$. If a
point-like instanton lives at a point $x\in S_H$ and the collision 
in the left of figure \ref{fig:sm} occurs at a point $y\in\Sigma_1$
then $\pi_H(x)=p(y)$. Thus by varying the complex structure of $X_1$
we may move the point-like instanton around in the ``base'' direction
of $S_H$.

In order to show that the moduli space of this point-like instanton is
given by $S_H$ we need to be able to vary its position in the
``fibre'' direction of $S_H$. This degree of freedom is provided by
the Ramond-Ramond degrees of freedom over $X_1$ as we now show.

On the right-hand side of figure \ref{fig:sm} the dotted line $f'$
denotes the proper transform of the $f$-curve which had passed through
the collision point on the left-hand side. 
As can be seen from the figure, $f'$ does not touch the discriminant
locus of the resulting elliptic fibration. Thus, the elliptic
fibration restricted to $f'$ has no bad fibres. This can only happen
if the fibre is constant. Let us refer to this constant elliptic curve
as $Q$. Thus $Q\times f'\subset X_1$.

Wedging by the generator of $H^2(f',\Z)$ gives an injective map
$H^1(Q,\Z)\to H^3(X_1,\Z)$. Thus $Q$ contributes 2 real degrees of
freedom to the Ramond-Ramond moduli.

We may rephrase this in terms of spectral curves again. As the \MW\
group of our generic rational elliptic surface is now trivial, one
might at first think that the spectral curve has simply collapsed to
the zero section of $\pi_H:S_H\to B$. Let us instead say that the
spectral curve, $C$, has degenerated in this case and has become
reducible, $C=B\cup Q$. 
Now the Jacobian of the spectral curve is isomorphic to the Jacobian of
$Q$. 
One can map this Jacobian into the intermediate Jacobian of $X_1$.
Again, this is essentially the Abel-Jacobi mapping of
\cite{Ty:Jac}. 

Now let us construct the moduli space of one instanton, $\cM_1$. As
argued above, there is a fibration $\cM_1\to B$ where the degree of
freedom in $B$ is generated by moving around the point of collision of
$\Delta$ within $C_0$. The fibre of this map is $\Jac(Q)$. However, since
$Q$ is constant all along $f'$, $Q$ is actually the elliptic fibre of
$\pi_H:S_H\to B$ over the same point in $B$. Thus we construct $\cM_1$
by replacing each elliptic fibre in $\pi_H:S_H\to B$ by its
Jacobian. Given that $\pi_H$ has 
a section, one may show that the resulting fibration has the same
complex structure as the original (for example, see proposition 5.3.2
of \cite{CDol:EnI}). {\em Thus we see that $\cM_1\cong S_H$ as
desired. }

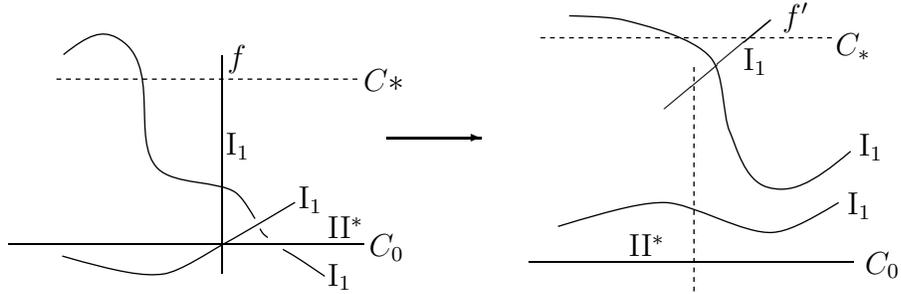
\begin{figure}
\begin{center}
\setlength{\unitlength}{0.00041700in}%
\begin{picture}(10887,3543)(889,-3673)
\thinlines
\put(901,-3061){\line( 1, 0){4500}}
\put(3601,-661){\line( 0,-1){2775}}
\put(5701,-1711){\vector( 1, 0){1200}}
\put(1501,-961){
}
\put(7501,-3286){\line( 1, 0){4125}}
\put(9601,-811){
}
\put(7651,-436){
}
\put(9226,-1336){\line( 6, 5){1350}}
\put(1576,-3211){
}
\put(7876,-2836){
}
\put(4379,-3128){
}
\put(1590,-653){
}
\put(8015,-158){
}
\put(4086,-2866){
}
\put(3676,-811){\makebox(0,0)[lb]{\smash{$f$}}}
\put(5476,-3211){\makebox(0,0)[lb]{\smash{$C_0$}}}
\put(4576,-2611){\makebox(0,0)[lb]{\smash{I$_1$}}}
\put(4951,-3586){\makebox(0,0)[lb]{\smash{I$_1$}}}
\put(4951,-2986){\makebox(0,0)[lb]{\smash{II$^*$}}}
\put(5401,-1111){\makebox(0,0)[lb]{\smash{$C*$}}}
\put(11701,-1936){\makebox(0,0)[lb]{\smash{I$_1$}}}
\put(11551,-2686){\makebox(0,0)[lb]{\smash{I$_1$}}}
\put(8776,-3211){\makebox(0,0)[lb]{\smash{II$^*$}}}
\put(11776,-3436){\makebox(0,0)[lb]{\smash{$C_0$}}}
\put(10726,-286){\makebox(0,0)[lb]{\smash{$f'$}}}
\put(11401,-661){\makebox(0,0)[lb]{\smash{$C_*$}}}
\put(3654,-1921){\makebox(0,0)[lb]{\smash{I$_1$}}}
\put(10231,-850){\makebox(0,0)[lb]{\smash{I$_1$}}}
\end{picture}
\end{center}
  \caption{The blow-up for a point-like $E_8$ instanton on a bad fibre.}
\label{fig:smI1}
\end{figure}

Actually this fibration of $\cM_1$ is a little more subtle than first meets the
eye. We have assumed that $Q$ is a smooth elliptic curve. That is, we
have assumed that our point-like instanton lives in a smooth fibre of
$\pi_H:S_H\to B$. It is quite possible however that it lives in a bad
fibre, such as an I$_1$ fibre. In this case the story of the
discriminant locus is given in figure \ref{fig:smI1}.

In this case $Q$ is a degenerate elliptic curve as one might expect. What is
interesting is how we now have a transverse intersection of two
curves of I$_1$ fibres in the discriminant. Such a collision produces
a singularity in $X_1$ which generically cannot be resolved. 
That is, {\em $X_1$ degenerates when we move the point-like instanton onto
a bad fibre of $S_H$.} This is the only way 
that we could have $\cM_1\cong S_H$. When we are on a bad fibre, the
intermediate Jacobian must degenerate and so $X_1$ must degenerate.

For more than one instanton we let
\begin{equation}
  C = B\cup \bigcup_{i=1}^{12-n} Q_i,
\end{equation}
where the $Q_i$ are the constant elliptic curves coming from the
scrolls formed by each of the blow-ups required. It is not hard to see
that our construction above will produce a Hilbert scheme of $12-n$ points
in $S_H$ as expected.

In the language above, our Prym is the product of the
Jacobians of $Q_i$ and so
\begin{equation}
  \Upsilon = \bigoplus_{i=1}^{12-n} H^1(Q_i,\Z).
\end{equation}

It is known that there is a close relationship between point-like
instantons on bad fibres and the very subtle case where the bundle on
$S_H$ in question is the tangent bundle. This has been discussed in
\cite{FMW:ell}. It was also explicitly seen in \cite{BCGJL:F} in terms
of a construction from toric geometry that the spectral
curve degenerated in precisely the way we have described here in the
tangent bundle case. 


\section{Discussion}   \label{s:disc}

Let $\Upsilon_1$ and
$\Upsilon_2$ be the lattices introduced in the last section from the
two $E_8$-bundles or sub-bundles thereof and let $M$ be the
transcendental lattice of the K3 on which the heterotic string is
compactified. Let $H^3(X,\Z)_0\cong H^3(X^\sharp,\Z)$ be the
monodromy-invariant part of  
$H^3(X,\Z)$ as we go around the stable degeneration discussed
in section \ref{s:sd}.

\def\BcM{B_{\!\!\cM}}
The boundary of the moduli space corresponding to this degeneration of
either the heterotic 
string or the type IIA string is of the following form.
It acquires
a fibration structure $\pi:\cM\to\BcM$ where the complex dimension of
$\BcM$ equals the quaternionic dimension of $\cM$ and the generic
fibre is an 
abelian variety $\C^{p}/L$ for some $p$ and some lattice
$L$. In identifying the moduli spaces, we must identify these abelian
varieties and thus the description of the lattice
$L$ in the heterotic picture and the type IIA picture. 

This leads to the key claim that
\begin{equation}
  H^3(X,\Z)_0 \cong M \oplus \Upsilon_1 \oplus \Upsilon_2.
		\label{eq:yes}
\end{equation}
This equation is essential to the notion of duality between the
heterotic string on a K3 surface and the type IIA string (or F-theory)
on a \CY\ threefold.

As well as automatically identifying the abelian fibres of the moduli
space, (\ref{eq:yes}) also indicates how to naturally identify the
bases, $\BcM$, as follows.%
\footnote{This argument only really shows how the Heterotic picture of
$\BcM$ and the type IIA picture of $\BcM$ can be identified
locally. It is possible that type IIA $\BcM$ may map many-to-one to
the heterotic $\BcM$. It is reasonable to expect this to happen when
the Prym has more than one component.}
 Recall that in the type IIA string
$\BcM$ represents the moduli space of complex structures on $X$ and
in the heterotic string $\BcM$ represents the moduli space of
complex structures on $S_H$ and the moduli of the spectral curve for
the bundle.
This map shows how the variation of Hodge structure on
$X$, in terms of periods of the holomorphic 3-form on elements of
$H^3(X,\Z)_0$, may be reduced into a statement about the variation of
Hodge structure on $S_H$, in terms of periods of the holomorphic
2-form on this K3 surface on elements of the transcendental lattice,
$M$. The bundle data  becomes encoded in a variation
of Hodge structure of a one dimensional object --- the spectral
curve. In the case of $\cG\cong\SU(N)$, $\Upsilon$ is simply
$H^1(C,\Z)$ and we recover the full moduli space of $C$. In general we
only have a sub-variation of Hodge structure for the spectral curve.

Thus the duality between the heterotic string and the type IIA string
appears to be encapsulated by relating the 3-dimensional structure of
the type IIA compactification to a 2-dimensional and a 1-dimensional
structure for the heterotic string. That is, the \CY\ threefold is
related to a K3 surface and a bundle.

It is interesting to compare this situation to mirror symmetry. One of
the simplest ways of describing the mirror symmetry principal for a
\CY\ threefold is the exchange of the even (or vertical) integral
cohomology with the odd (or horizontal) integral cohomology. This
picture was first suggested in the work of \cite{Cand:mir,AL:qag} and
then spelt out more clearly in section 5
of \cite{Mor:pred} and \cite{AM:Ud}.\footnote{Actually
this definition of mirror symmetry has yet to be shown to be
consistent with other definitions. In a few examples where the
integral cohomology is calculable and the rational curve can be
counted, it does appear to work.} 

The way that these integral
cohomologies can be identified requires one to go to the ``large
complex structure'' of one of the \CY\ threefolds. This was explained in
\cite{Mor:gid}. 
Only at this boundary {\em point\/} in the moduli space did one
expect the two mirror theories to be exactly equivalent. Away from
this boundary point the mirror equivalence becomes complicated by
world-sheet instanton effects which come from rational curves.

Heterotic/type IIA duality has a much richer structure than mirror
symmetry. Firstly our boundary where the two theories agree exactly is
no longer a point but has a fairly large dimension. Secondly one can
see that {\em both\/} the heterotic string and the type IIA string are
prone to instanton effects. Each of these can be probed individually
by moving away from the boundary in a certain direction.

For example, in order to probe the heterotic string world-sheet
instanton effects\footnote{There may also be higher-loop perturbative
corrections in $\alpha'$.} we need to move away from the boundary without
changing the dilaton in the type IIA string. In effect we need to keep
the section of $S_H$ infinitely large but we may let the elliptic
fibres become finite in size. Clearly this moves us away from the
stable degeneration discussed in section \ref{s:sd}. It is also
evident that the exact correspondence between our moduli spaces will
break down. Indeed we may go around a closed loop in the moduli space
of $X$ producing a highly non-trivial monodromy on $H^3(X,\Z)_0$.
If there were no quantum corrections to this moduli space then such a
monodromy would have to correspond to some sort of T-duality on the
moduli space of heterotic strings. This would imply that the moduli
space of complex structures on $X$ would be globally {\em exactly\/} in the
form of some Teichm\"uller space divided out by this T-duality
group. This is not true for a generic \CY\ threefold.
Thus the periods in our heterotic string on a K3 surface
must generically become 
``mixed'' between the variation of Hodge structure of $S_H$ and the
variation of Hodge structure of the spectral curve. That is, the notion
of a K3 surface and the notion of a bundle will become somewhat
confused. This is analogous to the way that the notion of a 0-cycle and
a 2-cycle can be confused by mirror symmetry away from the large
radius limit.

This confusion in the heterotic string is presumably accounted for by some
world-sheet instanton effect just as rational curves appeared in
mirror symmetry. It would be very interesting to study this further.

Another amusing fact in this picture is that we know that world-sheet
instanton effects must vanish for the heterotic string on a K3 surface
if ``the spin connection if embedded in the gauge group''. In other
words, when our bundle becomes the tangent bundle. If this can be
understood in the language of this paper then it should be quite a
potent weapon in understanding the global structure of the moduli
space further. It implies that there is a whole $\GO(\Gamma_{4,20})
\backslash\GO(4,20)/(\GO(4)\times\GO(20))$ exact subspace of the moduli
space coming from deformations of the K3 surface. This probes deeply
into regions of the moduli space where the type IIA string becomes
strongly coupled. Thus we may calculate instanton effects on the type
IIA side too.

We should emphasize that there is a big difference between mirror
symmetry and the map (\ref{eq:yes}). The map given in (\ref{eq:yes})
does not include the whole of $H^3(X,\Z)$ and it does not contain the
whole of $H^2(S_H,\Z)$. In fact given the nature of the moduli space
of strings on K3 surfaces \cite{AM:K3p} one might hope that such a map
should include the total cohomology $H^*(S_H,\Z)$. It would be very
satisfying, as well as useful, if we could enlarge the identification 
of (\ref{eq:yes}) to include these larger integral structures.

In conclusion, the moduli space of hypermultiplets of $N=2$ theories
in four dimensions coming from compactifications of the type IIA
string or the heterotic string contains some beautiful
structures. These structures are similar to those of mirror symmetry
but appear to be much more powerful. As well as providing insight into
the nonperturbative properties of $N=2$ theories in four dimensions we
may also learn some lessons about nonperturbative effects in general
in string theory.


\section*{Acknowledgements}

Much of the earlier stages of this work were done in collaboration
with D.~Morrison. I am also particularly grateful to R.~Donagi and
R.~Hain for explaining the mathematics of the appendix to me.
It is a pleasure to thank M.~Gross, D.~Reed
and E.~Witten for useful conversations. 


\section*{Appendix: Dimension Formula from Pryms%
\footnote{This section was explained to me by R.~Donagi and R.~Hain.}}

We wish to prove the following:
\begin{theorem}
Let $W$ be the Weyl group of some given Lie group $\cG$ and let
$\Lambda$ be the root lattice of $\cG$. Let 
$\pi:\widetilde C\to B$ be a $W$-Galois Cameral cover,
$\widetilde C$ be irreducible and $B$ be a curve of genus 0. Then in
the generic case 
\begin{equation}
\begin{split}
  l &= \ff12\dim\Hom_W(\Lambda,H^1(\widetilde C,\C))\\
    &= \ff12b-\rank(\cG),
\end{split}
\end{equation}
where $b$ is the number of branch points.
\end{theorem}

Consider the {\em representation ring\/} of $W$ denoted $R(W)$. An
element of this ring is a formal sum of the form
\begin{equation}
  \sum n_\alpha[V_\alpha],
\end{equation}
where $n_\alpha\in\Q$ and $V_\alpha$ is an irreducible
representation. There is an augmentation map
\begin{equation}
  \varepsilon:R(W)\to\Z,
\end{equation}
given by $\varepsilon([V_\alpha])=\dim(V_\alpha)$.

Let us denote the simplicial chain complex of $\widetilde C$ by
$\widetilde S_\bullet$ and the simplicial chain complex of $B$ by
$S_\bullet$. We may regard $\widetilde S_\bullet$ and $S_\bullet$ as
$\C(W)$-modules (where the action of $W$ on $S_\bullet$ is trivial).

The fibre of the covering $\pi$ is a set of points
corresponding to the Weyl chambers of $W$ except over the points
$p_j\in B$, $j=1,\ldots,b$, where the covering is branched. 
At each
$p_j$ there is a 
monodromy action on the fibres given by an element of the Weyl group
which we denote $\sigma_j\in W$.

It follows that, as $\C(W)$-modules,
\begin{equation}
\begin{split}
  \widetilde S_k &= S_k \otimes_{\C} \C(W) \qquad\hbox{for $k>0$}\\
  \widetilde S_0 &= \left(S_0-\bigcup_{i=1}^b\{p_j\}\right)
        \otimes_{\C}\C(W) \oplus\left(
        \bigoplus_{j=1}^b\{p_j\}\otimes_{\C}\C(W/\langle\sigma_j\rangle)
	\right).
\end{split}
\end{equation}

One may show from standard methods that the Euler characteristic of
the curve $\widetilde C$ is equal to the Euler characteristic of the
chain complex $\widetilde S_\bullet$. Regarding $\widetilde S_\bullet$
as a chain complex of $\C(W)$-modules we may define $\hat\chi
\in R(W)$ as a refinement of the usual Euler characteristic. The
object $\hat\chi$ reduces to the usual Euler characteristic under the
map $\varepsilon$. By the same argument as above $\hat\chi(\widetilde
C)=\hat\chi(\widetilde S_\bullet)$. Thus
\begin{equation}
  \hat\chi(\widetilde C) = \chi(B)[\C(W)] - b[\C(W)]
      + \sum_{j=1}^b\left[\C(W/\langle\sigma_j\rangle)\right].
\end{equation}

It is now a straight-forward matter of using the representation theory
of groups to determine $l$. Firstly one can show that 
$\dim\Hom_W(\Lambda,\C(W))=\rank(\cG)$. The statement that our Cameral
cover is generic amounts to the statement that the monodromy about any
$p_j$ corresponds to a simple Weyl reflection in a single root.
Then $\dim\Hom_W(\Lambda,\C(W/\langle\sigma_j\rangle))$ is the
dimension of the root space of $\cG$ invariant under such a reflection
which is $\rank(\cG)-1$.

We also know that $H^0(\widetilde C)$ and $H^2(\widetilde C)$ live in
the identity representation of $W$. Thus $\dim\Hom_W(\Lambda,H^k(
\widetilde C))=0$ for $k=0$ or $k=2$. It follows that
\begin{equation}
\begin{split}
  \dim\Hom_W(\Lambda,H^1(\widetilde C,\C)) &= -b(\rank(\cG)-1) - 
	(2-b)\rank(\cG)\\
  &= b - 2\rank(\cG). \qquad\qquad\qquad\qquad\qquad\qquad\square
\end{split}
\end{equation}



\end{document}